\documentclass[aps,showpacs,nofootinbib,prd,twocolumn]{revtex4-1}
\usepackage{graphicx}
\usepackage{amssymb}
\usepackage{amsmath}
\usepackage[svgnames]{xcolor}
\usepackage{mathtools,slashed}
\usepackage[retainorgcmds]{IEEEtrantools}
\usepackage{epstopdf}
\usepackage[utf8]{inputenc}
\usepackage{url}
\usepackage[colorlinks,citecolor=DarkGreen,linkcolor=DarkRed,urlcolor=DarkBlue]{hyperref}
\usepackage[compat=1.1.0]{tikz-feynhand}
\usepackage{tikz}
\usetikzlibrary{arrows.meta}
\usepackage{newtxtext}
\usepackage{newtxmath}

\usepackage{epsfig}

\newcommand{\be}{\begin{eqnarray}}
\newcommand{\ee}{\end{eqnarray}}
\begin{document}

\title{Estimate for the neutrino magnetic moment from pulsar kick velocities induced at the birth of strange quark matter neutron stars}

\author{Alejandro Ayala$^1$}
\email{ayala@nucleares.unam.mx}
\author{Santiago Bernal-Langarica$^1$}
\email{santiago.bernal@correo.nucleares.unam.mx}
\author{Daryel Manreza-Paret$^2$}
\email{dmanreza@gmail.com}

  \address{
  $^1$Instituto de Ciencias
  Nucleares, Universidad Nacional Aut\'onoma de M\'exico, Apartado
  Postal 70-543, CdMx 04510,
  Mexico.\\
  $^{2}$Facultad de Física, Universidad de La Habana, San Lázaro y L, La Habana, Cuba.
}

\begin{abstract}

We estimate the magnetic moment of electron neutrinos by computing the neutrino chirality flip rate that can occur in the core of a strange quark matter neutron star at birth. We show that this process allows neutrinos to anisotropically escape, thus inducing the star kick velocity. 
Although the flip from left- to right-handed neutrinos is assumed to happen in equilibrium, the no-go theorem does not apply because right-handed neutrinos do not interact with matter and the reverse process does not happen, producing the loss of detailed balance. For simplicity, we model the star core as consisting of strange quark matter. We find that even when the energy released in right-handed neutrinos is a small fraction of the total energy released in left-handed neutrinos, the process describes kick velocities for natal conditions, which are consistent with the observed ones and span the correct range of radii, temperatures and chemical potentials for typical magnetic field intensities. The neutrino magnetic moment is estimated to be $\mu_\nu \sim 3.6 \times 10^{-18}\mu_B$, where $\mu_B$ is the Bohr magneton. This value is more stringent than the bound found for massive neutrinos in a minimal extension of the \mbox{standard model.}
\end{abstract}
\maketitle


\section{Introduction}

Compact objects are a class of astrophysical objects that include White Dwarfs (WD), Neutron Stars (NS), and Black Holes (BH). A~significant proportion of their physical properties remain unknown since their initial appearance in the astronomical landscape. These types of objects are particularly interesting because many areas of physics have converged into their~study. 

One of the main aspects that have been studied in recent years is the internal composition of WD and NS, due to the extremely high densities, temperatures, and magnetic fields that are found in these systems, which terrestrial laboratories~\cite{MPD:2022qhn} have little chance of reproducing. 
One of the aims of studying these kinds of systems is to probe the not yet understood phases of strongly interacting matter~\cite{Gutierrez:2013sta,Pasztor:2024dpv,Du:2024wjm} in the regime of high density, low temperature, and large magnetic fields~\cite{Radice:2024gic}, which determine the proposed scenarios of the exotic phases of matter that compose the internal layers of NS~\cite{Lugones:2024ryz,Issifu:2024fuw,Schramm:2011aa}.

NS are born after a massive star, with a mass larger than 8 solar masses, explodes as a Type II supernova~\cite{Burrows:2020qrp,Boccioli:2024abp}. It has been realised that the properties of NS are related to the continuous neutrino emission during the first 100 years after NS formation, which rapidly cools down the star from dozens of MeV to {only  dozens of keV}. If~the emission of neutrinos takes place in an anisotropic way, it can be responsible for the so-called pulsar kicks~\cite{Hobbs:2005yx}.

Several scenarios have been proposed to explain the proper motion of pulsars~\cite{Fukushima:2024cpg,Berdermann:2006rk,Janka:2016nak}, including hydrodynamic perturbations~\cite{Gessner:2018ekd,Nordhaus:2010ub,Page:2020gsx}, asymmetric electromagnetic emissions from an off-centered magnetic dipole~\cite{1975Natur.254..676T,1975ApJ...201..447H,Lai:2000pk,Agalianou:2023lvv}, {the breaking of a binary system~\cite{1970ApJ...160L..91G,Igoshev:2021bxr},} and anisotropic neutrino emission~\cite{Wongwathanarat:2012zp,Sagert:2007as,Fryer:2005sz}, among~others~\cite{Wang:2005jg,Khokhlov:1999ka,Colpi:2002cu,Fryer:2003jj,Farzan:2005yp,Lambiase:2023hpq}. One of the main scenarios is a natal kick, which takes place during the formation of the proto-NS, in~which neutrinos are emitted in a preferred direction and, hence, provide the proto-NS with momentum in its opposite direction. This scenario is favored by Refs.~\cite{Janka:2017vlw,Sieverding:2019qet}, where it has been shown that there is a peak in the electron neutrino luminosity of about 12 s after the core bounce stage of proto-NS formation. It has also been proposed in Refs.~\cite{Sumiyoshi:2017fpf,Keranen:2004vj} that the formation of quark matter during the core collapse could explain the observed NS~kicks.

In a minimal extension of the standard model (SM), where neutrinos are massive and can have a non-vanishing magnetic moment~\cite{Fujikawa:1980yx,Alok:2022pdn,Alok:2023sfr}, we recently computed the NS kick velocities by considering the case in which the anisotropic neutrino emission has its origins in the presence of a strong magnetic field in the interior of an NS~\cite{Ayala:2018kie}. If~the neutrinos, due to their magnetic moment, interact in equilibrium with the NS medium, they can flip their chirality, becoming right-handed and, hence, suppressing their interactions with matter. As the inverse process happens out of equilibrium and the detailed balance is lost, the~right-handed neutrinos cannot flip back onto the left-handed ones, making the no-go theorem non-applicable~\cite{Kusenko:1998yy}. Furthermore, because~matter effects dominate in the core of NS, the~right-handed neutrinos cannot resonate back into left-handed ones, as this process is suppressed as a consequence of the presence of magnetic fields. A~fraction of the emitted left-handed neutrinos can escape from NS as right-handed ones if the typical time that takes them to flip their chirality is smaller than the time needed to travel one mean free path, which is small compared with the NS core radius. This mechanism was implemented in Refs.~\cite{Ayala:2019sbt,Ayala:2021nhx} and was used to establish a lower bound for the neutrino magnetic moment that, together with the most stringent upper bound for this magnetic moment~\cite{Ayala:1998qz,Ayala:1999xn}, allowed for setting the range $4.7 \times 10^{-15} \leq \mu_\nu/\mu_B \leq$ (0.1--0.4) $\times$ $10^{-11}$.  In~this work, we complement that study by further elaborating on the idea that a neutrino chirality flip, produced by the existence of a neutrino magnetic moment, can explain the observed kick velocities. As SN explosion is mainly driven by neutrino emission, not all of the neutrinos should flip their chirality; ~otherwise, the explosion itself would not take place. We show that even if a small percentage of neutrinos become right-handed from the original left-handed state, the~observed kick velocities can be reproduced even for a neutrino magnetic moment of the order of SM bound for a neutrino mass of a few eV~\cite{Fujikawa:1980yx}.

\section{Production Rate of Right-Handed~Neutrinos}

The chirality flip within the core of NS can be considered in the same way as it was discussed in Refs.~\cite{Ayala:1998qz,Ayala:1999xn}, ion order to place an upper bound on the neutrino magnetic moment from the chirality flip in supernovae. To~this end, we consider a thermally equilibrated plasma at a temperature $T$ and electron chemical potential $\mu_e \gg m_e$ in which neutrinos are being produced, with~$m_e$ being the electron mass. For~the sake of simplicity, throughout this work, we consider the neutrinos as massless. The~production rate of a right-handed neutrino, from~a left-handed one, whose energy is $p_0$ and its momentum is $\vec{p}$, is given by
\begin{equation}
\label{neuchir1}
    \Gamma (p_0) = \frac{\tilde{f}(p_0)}{2p_0} \text{Tr}\left[\slashed{P}R\text{Im}\Sigma\right],
\end{equation}
where $\tilde{f}(p_0)$ is the Fermi--Dirac distribution for right-handed neutrinos, such that $P_\mu = (p_0,\vec{p})$, $|\vec{p}| = p$ {, where we use the Feynman slash notation $\slashed{a} = \gamma^\mu a_\mu$, with~$\gamma^\mu$ as the Dirac matrices}. In~the previous equation, the~operators {$R,\: L= \frac{1}{2}(1\pm\gamma_5)$ projects onto right (left)-handed} fermion components{, where the chiral matrix $\gamma^5$ is defined as $\gamma^5 \equiv i\gamma^0\gamma^1 \gamma^2 \gamma^3$}. The~neutrino self-energy $\Sigma$ is given by
\begin{equation}
\label{neuchir2}
\Sigma(P)=T\sum_n \int \frac{d^3 k}{(2\pi)^3} V ^\rho (K) \, S_F(\slashed{P} - \slashed{K})\, L \, V ^\lambda (K) \, D_{\rho\lambda} (K),
\end{equation}
where $K_\alpha = (k_0,\vec{k})$, $|\vec{k}|=k$, $V^\mu$ is the neutrino--photon vertex function, $S_F$ is the neutrino propagator and $D_{\rho\lambda}$ is the photon propagator. 
We describe the neutrino--photon interaction by means of a magnetic dipole interaction $V _\mu(K) = \mu_\nu \, \sigma_{\alpha\mu}K^\alpha$, where $\mu_\nu$ is the neutrino magnetic moment and $\sigma_{\alpha\mu} = \frac{i}{2} \left[\gamma_\alpha,\gamma_\mu\right]$. As~usual, we split the photon propagator into its longitudinal and transverse components
\begin{equation}
\label{neuchir3}
D_{\rho\lambda} (K) = \Delta_L (K) P^L _{\rho \lambda} + \Delta _T (K) P^T _{\rho \lambda}.
\end{equation}

The sums over Matsubara frequencies that need to be evaluated are
\begin{equation}
M_{L,T}=T\sum_n \Delta_{L,T} (i\omega_n)\tilde{\Delta}_F (i(\omega-\omega_n)).
\end{equation}

By writing the neutrino and photon propagators in their spectral representation, the~imaginary part of $M_{L,T}$ can be computed in a straightforward manner,~as follows:
\begin{widetext}
\begin{IEEEeqnarray}{rCl}
\mathrm{Im}\left[M_{L,T}\right] & = & \pi\left( e^{\beta (p_0 - \mu)} + 1 \right) \int_{-\infty} ^\infty \frac{dk_0}{2\pi} \int_{-\infty} ^\infty \frac{dp'_0}{2\pi} f(k_0)\tilde{f}(p'_0 - \mu) \delta(p_0 - k_0 - p'_0) \rho_{L,T} (k_0,k) \rho_F (p'_0),
\label{neuchir6}
\end{IEEEeqnarray}
where $f(k_0)$ is the Bose--Einstein distribution, and $\rho_{L,T}$ and $\rho_{F}$ are the spectral densities for the photon and neutrino propagators, respectively, given by
\begin{IEEEeqnarray}{rCl}
\label{neuchir7}
\rho_F(p'_0) & = & 2\pi \delta({p'_0} ^2 - E_p ^2)\: \text{sign} (p'_0),\nonumber\\
\rho_L(k_0,k) & = & \frac{x}{1-x^2}\frac{2\pi m_\gamma^2\theta(k^2-k_0^2)}{\left[k^2+2m_\gamma^2\left(1-\frac{x}{2}\ln |(1+x)/(1-x)|\right)\right]^2+\left[\pi m_\gamma^2x\right]^2} ,
\nonumber\\
\rho_T(k_0,k) & = & \frac{\pi m_\gamma^2 x(1-x^2)\theta(k^2-k_0^2)}{\left[k^2(1-x^2) + m_\gamma^2\left(x^2+\frac{x}{2}(1-x^2)\ln \left|(1+x)/(1-x)\right|\right)\right]^2  + \left[(\pi/2)m_\gamma^2x(1-x^2)\right]^2},\IEEEeqnarraynumspace
\end{IEEEeqnarray}
where we defined= $x =k_0/k$ , and  $\text{sign}(p'_0) $ is the sign function{, $\theta$ is the Heaviside function},
and the photon thermal mass $m_\gamma$ is given by
\begin{equation}
  m_\gamma^2 = \frac{e^2}{2\pi^2}\left(\mu_e^2 + \frac{\pi^2 T^2}{3}\right).
\label{photonnmass}
\end{equation}

Using Equation~\eqref{neuchir6}, one finds that the expression for $\Gamma$ involves the factor
\begin{IEEEeqnarray}{C}
    \label{neuchir9}
 \pi\left( e^{\beta (p_0 - \mu)} + 1 \right)\int \frac{d^3 k}{(2\pi)^3}\int_{-\infty} ^\infty \frac{dk_0}{2\pi} f(k_0) \left(\sum_{i=L,T}C_i \rho_i (k_0)\right) \nonumber \\
\times \:\frac{1}{2E_p}\left[ \tilde{f}(E_p - \mu)\delta(E_p - (p_0-k_0))   - \left(1-\tilde{f}(E_p + \mu)\right)\delta(E_p - (p_0-k_0)) \right],
\end{IEEEeqnarray}
where the contraction of the traces with the longitudinal and transverse polarization tensors give rise to the functions $C_L$ and $C_T$, whose explicit expression is
\begin{IEEEeqnarray}{rCl}
    C_L &=& -k^2 \left(1-x^2\right)^2 \left(2p_0 - k_0\right)^2 ,\\
    C_T &=& k^2 \left(1-x^2\right)^2 \left[\left(2p_0 - k_0\right)^2 - k^2\right].
\end{IEEEeqnarray}

We can identify two kinds of terms in Equation~\eqref{neuchir9}: the term that corresponds to a left-handed neutrino in the initial state, which is proportional to $\tilde{f}(E_p - \mu)$, and~the term that corresponds to the left-handed neutrino in the final state, which is proportional to $\left(1-\tilde{f}(E_p + \mu)\right)$. Given that we are looking to produce right-handed neutrinos from left-handed ones, only the contribution of the term proportional to $\tilde{f}(E_p - \mu)$ must be taken into account. By~performing the angular integrals, we arrive at an expression for the production rate of right-handed neutrinos as a function of $p_0$
\begin{IEEEeqnarray}{rCl}
    \label{neuchir10}
    \Gamma(p_0)&=&  \frac{\mu_\nu ^2}{32\pi^2 p_0 ^2} \int_0 ^\infty dk \, k^3 \int_{-k} ^{k} dk_0\ \theta(2p_0 +k_0 - k) [1+f(k_0)]\, \tilde{f}(p_0 + k_0 -\mu) \, (2p_0 + k_0)^2 \left(1-\frac{k_0^2}{k^2}\right)^2 \nonumber \\
    & & \times \: \left[ \rho_L (k_0) + \left( 1 - \frac{k^2}{(2p_0 + k_0)^2}\right) \rho_T (k_0) \right].
\end{IEEEeqnarray}

{Notice that, even when neglecting the neutrino mass, one could in principle consider that in a large density scenario, the~neutrino could develop a Debye mass~\cite{PerezMartinez:1996ms}. However, this mass turns out to be of the order of keV, which is small compared with the Debye mass developed by photons that, as~we show below, for~typical conditions of an NS at birth, turns out to be of an order of a couple dozens of MeV. Therefore, hereafter, we also neglect a possible neutrino Debye mass.} 

\end{widetext}

The total reaction rate is obtained as the integral of $\Gamma$ over the available phase space, that is,
\begin{equation}
    \Gamma= V\int\frac{d^3p}{(2\pi)^3}\Gamma(p_0)
    =\frac{V}{2\pi^2}\int_0^{p_0^{max}}dp_0\ p_0^2\ \Gamma(p_0).
    \label{totalrate}
\end{equation}
$V$ represents the volume in which the chirality flip process occurs and the upper limit of the integral over the energy is determined by the maximum energy allowed for the neutrino in the beta decay process. We fix $p_0 ^{max} =1.2$ MeV, which represents the maximum value for the energy of the massless neutrino beta decay in a vacuum. This is due to the fact that the process is mediated by a $W$ boson, whose mass is significantly larger than the typical energy scales that are present during the formation of a~proto-NS.

\section{Pulsar Kick~Velocity}

The resulting kick velocity for NS can be calculated using the following expression~\cite{Sagert:2007as,Sagert:2007ug}
\begin{equation}\label{vel_kick}
dv=\frac{\chi }{M_{NS}}\frac{4}{3}\pi R^3\epsilon dt,
\end{equation}

$\chi$ is the electron spin polarization (a function only of temperature, density, and the magnetic field intensity after the neutrinos flip their chirality); 
$\epsilon$ is the neutrino emissivity; and $M_{NS}$ and $R$ are the  mass and radius, respectively, of the NS~core.
 
{As the neutrinos are predominantly emitted along the direction opposite the magnetic field, the~kick velocity is naturally aligned with the direction of the magnetic field. This seems in line with recent findings of a correlation, that is built along time, which also exists between the direction of the magnetic field and the NS angular velocity~\cite{Young:2009iy,2021NatAs...5..788Y,Noutsos:2012dt}.}

We point out that as NS are compact and massive, the~emission of energy from these objects in any form can receive corrections from general relativistic (GR) \mbox{effects~\cite{Pons:1998mm, Page:2005fq, Potekhin:2015qsa}}. 
In particular, the~neutrino luminosity is proportional to the metric factor $e^{2\Phi} \approx 0.5$. Hence, the~GR effects reduce the neutrino luminosity by about half. Nonetheless, the~asymmetry in the emission is not affected by GR effects, and the maximum energy released as neutrinos is still about $10^{53}$ erg. We expect that a small fraction of this energy is emitted in the form of right-handed neutrinos and, later on, we provide an estimate of this fraction. For~a discussion on other GR effects, see Ref.~\cite{Schmitt:2017efp}.

The right-handed neutrino emissivity $\epsilon$ can be expressed in terms of the left-handed neutrino emissivity $\varepsilon$ as
\begin{equation}
    \epsilon = g \tau \Gamma \varepsilon,
    \label{rhnemissivity}
\end{equation}
where $\Gamma$ is the total reaction rate for the chirality flip, $\tau$ is the emission time-scale, and $g$ is a factor that accounts for the fraction of total energy carried away by the right-handed neutrinos. When the emissivity changes with temperature, the~cooling equation can be used, namely,
\begin{equation}\label{vel_kick3}
-\epsilon=g\tau\Gamma\frac{dU}{dt}=g\tau\Gamma\frac{dU}{dT}\frac{dT}{dt}=g\tau\Gamma C_v\frac{dT}{dt},
\end{equation}
where $U$ is the internal energy density and  $C_v$ is the heat capacity. Therefore, the~kick velocity is given by
\begin{equation}\label{velchi1}
v=-\frac{g}{M_{NS}}\frac{4}{3}\pi R^3\tau\int_{T_i}^{T_f}\Gamma\chi\, C_vdT.
\end{equation}

This velocity can be written in the following form
\begin{equation}
v=-g\ {804} \ \frac{\text{km}}{\text{s}}\left( \frac{1.4M_\odot}{M_{NS}}\right)\left( \frac{R}{10 \ \text{km}}\right)^3 \left( \frac{\tau I}{\text{MeV} \ \text{fm}^{-3}}\right),
\label{velchi2}
\end{equation}
where
\begin{equation}\label{vel2}
I=\int_{T_i}^{T_f}\Gamma\,\chi\, C_vdT.
\end{equation}

The latter integral depends on $\Gamma$, $C_v$, and $\chi$, which, in turn, are given in terms of $B$, $\mu_i$, and $T$. The~electron spin polarization and the heat capacity were derived and studied in Ref.~\cite{Ayala:2018kie}.

\begin{widetext}
The particle density can be computed as
\begin{equation}
    n_f = \displaystyle{\frac{d_f m_f ^3}{2\pi^2}b_f\sum_{l=0}^\infty (2-\delta_{l0}) \int_0 ^\infty dx_3 \,\frac{1}{\exp{\frac{m_f}{T}\left(\sqrt{x_3^2 + 1 +2lb_f} - x_f\right)}+1}},
\end{equation}
where $m_f$ is the fermion mass, $l$ is the Landau level, and the~dimensionless momentum is $x_3=p_3/m_f$  with $p_3$ being the momentum along the magnetic field direction $\textbf{B}$, and~$b_f=B/B_f^c$, with~$B_f^c=m_f^2/e_f$ being the critical magnetic field. We use the dimesionless quantity  $b_e\equiv b=B/B_e^c$. The~heat capacity is given by $C_{v}=\sum_f C_{vf}$, where
\begin{equation}\label{Cv}
C_{vf}=\frac{d_fm^2}{4\pi^2 T^2}b_f\int_{0}^\infty dp_3\sum_{l=0}^{\infty}(2-\delta_{l0})\frac{(E_{lf}-\mu_f)^2}{[1+\cosh{\frac{E_{lf}-\mu_f}{T}}]},
\end{equation}
and $E_{lf}$ is given by
\begin{equation}\label{Disp-Rel}
E_{lf}=(p_3^2+2|e_f|Bl+m_f^2)^{1/2},
\end{equation}
with $e_f$ being the fermion charge. The~electron spin polarization is given by
\begin{equation}\label{chi2}
\chi=\left\lbrace 1+ \frac{2
\sum\limits_{\nu=1}^{\infty}\int_0^\infty dx_3 \frac{1}{e^{\frac{m_e}{T}\left(\sqrt{x_3^2+1+2\nu b}-x_e\right)}+1}}{\int_0^\infty dx_3 \frac{1}{e^{\frac{m_e}{T}\left(\sqrt{x_3^2+1}-x_e\right)}+1}} \right\rbrace ^{-1},
\end{equation}
where $x_3=p_3/m_f$ is the dimensionless~momentum.

For small field strengths, the sum over Landau levels appearing in the particle density, heat capacity, and electron spin polarization expressions can be approximated by resorting to the Euler--McLaurin formula~\cite{Ayala:2021nhx}
\begin{equation}
    h\left[\frac{f_0}{2}+f_1+\ldots +\frac{f_N}{2}\right] = \int_0 ^{Nh} dx\, f(x) + \sum_k \frac{B_{k+1}}{(k+1)!}h^{k+1} (f_N ^{(k)} - f_0 ^{(k)}),
\end{equation}
where $B_k $ are the Bernoulli numbers. By~assuming the limit $N\to\infty$ and keeping terms up to order $h^2$, we can approximate the expressions for $n_f$, $C_{vf}$ and $\chi$ as 
\begin{IEEEeqnarray}{rCl}
    n_f & = &\frac{d_f m_f^3}{2\pi^2} \left[\int_0 ^\infty dx \, \frac{x^2 + b_f}{\exp(\frac{m_f}{t}\left(\sqrt{x^2+1}-x_f\right))+1} \right. + \left. \frac{b_f ^2 m_f}{24 T}\int_0 ^\infty dx \frac{\text{sech}^2\left(\tfrac{m_f}{T}\left(\sqrt{x^2+1} - x_f\right)\right)}{\sqrt{x^2+1}}\right],  \label{eq:density}\\
    C_{vf} & = & \frac{d_f}{4\pi^2 t^2} \left\{\frac{m_f ^5}{|e_f|B_f ^c}\int_0 ^\infty dx\, \frac{\left(x^2 - \tfrac{b_f}{m_f ^2} |e_f|B_f ^c\right) \left(\sqrt{x ^2 + 1} - x_f\right)^2}{1+\cosh(\frac{\sqrt{x^2+1} - x_f}{t})} + \frac{|e_f| B_f ^c b_f ^2 m_f}{3}  \right. \nonumber \\
    & \times & \int_0 ^\infty dx\,\frac{\left(\sqrt{x^2 + 1} - x_f\right)}{\sqrt{x^2 + 1}} \left.  \frac{\left[-1 - \cosh(\tfrac{\sqrt{x^2 + 1} - x_f}{t}) + \frac{\sqrt{x^2 + 1}-x_f}{2t}\sinh(\frac{\sqrt{x^2 + 1}-x_f}{t})\right]}{\left(1+\cosh(\frac{\sqrt{x^2+1}-x_f}{t})^2\right)}\right\},\label{eq:Cv} \IEEEeqnarraynumspace\\
    \chi^{-1} & = & 1 + \frac{2\left( \int_1 ^\infty dx\, \frac{x^2}{\exp(\frac{m_e}{t}\sqrt{x^2 + 1} - x_e) + 1} + \frac{b^2 m_e}{24 T}\int_0 ^\infty dx\, \frac{\sec^2\left( \frac{m_e}{T}\left( \sqrt{x^2 + 1} - x_e\right) \right)}{\sqrt{x^2 + 1}}\right)}{\int_0 ^\infty dx\, \frac{1}{\exp(\frac{m_e}{T}\sqrt{x^2+1}-x_e) +1}}  , \label{eq:chi}
    \label{eq:EulerMcLaurinApprox}
\end{IEEEeqnarray}
where $x$ is a dimensionless variable and we use $h = 2b_f$ for $n_f$, $h = 2|e_f|B_f ^c b_f$ for $C_{vf}$ and $h = 2b$ for $\chi$.
\end{widetext}

For simplicity, we consider the core of NS as a plasma made out of magnetized strange quark matter (SQM), namely, a~gas composed of quarks $u$, $d$, and $s$ and electrons $e$ in the presence of a magnetic field. SQM may exist in the core of some heavy neutron stars~\cite{Annala:2019puf}, and~we take this type of matter as the framework for our calculations. 
The formation of SQM in the interiors of NS can happen if the whole star makes a transition into SQM~\cite{Pagliara:2013tza,Witten:1984rs}. Nonetheless, the~existence of hybrid stars, in~which a sharp interface between hadronic and quark matter phases or a mixed phase region can be found, have also been proposed~\cite{Alford:2004pf,Schramm:2011aa,Schramm:2019oub}. In~this work, we considered a hybrid situation in which SQM only exists in the core of NS.
{Notice that, as~pointed out, for instance in Ref.~\cite{Annala:2019puf}, the~largest mass NS, with~masses around twice the solar mass, have characteristics consistent with a phase of quark matter in the core. Nevertheless, the~mass-radius relation of other NS, with~masses around 1.4 solar masses, seems to be reproduced by EoS and is compatible with an SQM core. 
	 Hence, although~our calculation is applicable to NS with an SQM core, this does not imply that it is only applicable to the largest mass stars nor that the fastest are the largest mass ones, as other parameters, such as temperature and magnetic field strength, also play a role to determine the kick velocity.}

During the first seconds of the evolution of NS, the~three quark species present in the core are in $\beta$ equilibrium; among them, the~core must maintain a neutral charge, and there should be a conservation of the baryon number and a fixed fraction of electrons plus neutrinos to baryons of $Y_L = 0.4$~\cite{Prakash:1995uw,Reddy:1997yr}. 
These conditions are referred to as stellar equilibrium~conditions.

\section{Bound for the Neutrino Magnetic~Moment}

From Equations~\eqref{totalrate} and \eqref{eq:density}--\eqref{eq:chi}, we can compute the value of the induced kick velocity. Notice that  this velocity is determined by the neutrino magnetic moment. Therefore, we can use typical values of NS kick velocities, together with average values of NS mass, radii, magnetic field intensities, temperature, and chemical potentials at birth, to~find an estimate for the neutrino magnetic moment. To~this end, we assume that the volume where the chirality flip takes place is a cylinder whose height is a neutrino mean-free path, $\lambda\sim 1$ cm~\cite{Ouyed:2022toz}, and the transverse size is determined by the radius $R$ of NS.
The process needs to take place within the first 10--30 s after the core collapse; otherwise, the left-handed neutrino mean free path becomes larger than the radius of the star and the neutrinos escape, producing the SN explosion.
Therefore, for~$R\sim 10$ km, the~total volume where the reaction takes place is
\begin{equation}
    V = \pi R^2 \lambda \approx 3\times 10^{51} \text{ fm}^3 .
\end{equation}

Also, the~time interval where the flip takes place is taken as the time for the neutrinos to travel one mean free path at the speed of light, $\tau = \lambda/c \sim 3\times 10^{-11}$ s. We assume that, at birth, the~temperature ranges between $T = 30$--50 MeV, the~neutrino chemical potential is $\mu \sim 300$ MeV, and the electron chemical potential is $\mu_e \sim 275$ MeV. We estimate the fraction of the total energy carried away by the righ-handed neutrinos $g$ as the fraction of the product of the volume times the emission time where the flip takes place $V\tau\sim$~$9$~$\times 10^{40}$~fm$^3$~s, to~the total NS core volume times the total neutrino emission time $V_{NS} \tau_{NS}$~$\sim$~1.2~$\times$~$10^{59}$~fm$^3$~s, giving
\begin{eqnarray}
    g\sim 7.5 \times 10^{-19},
\end{eqnarray}
which means that, if~the energy released in the left-handed neutrinos is around $10^{53}$ erg, then the energy released in the right-handed neutrinos is around $10^{34}$ erg. Using these estimates, we find that the NS kick velocity is
\begin{equation}
    v \sim 3.1 \times 10^{37} \left(\frac{\mu_\nu}{\mu_B}\right)^2 \text{ km s}^{-1}.
\end{equation}

Recall that the average observed birth velocity for pulsars whose characteristic age is less than three million years is $v\approx 400 \text{km s}^{-1}$~\cite{Hobbs:2005yx}. Using this value, we estimate that the neutrino magnetic moment is of the order
\begin{equation}
    \mu_\nu \sim 3.6 \times 10^{-18}\mu_B.
\end{equation}

On the other hand, the~minimal extension of the SM expression for the neutrino magnetic moment, for~neutrino with mass $m_\nu$, at~the one-loop level, is given by~\cite{Fujikawa:1980yx}
\begin{equation}
    \mu_\nu = \frac{3eG_F m_\nu}{8\pi^2\sqrt{2}} = \frac{3G_F m_e m_\nu}{4\pi^2\sqrt{2}}\mu_B.
\end{equation}

By considering the parameters obtained from the solar, atmospheric, and reactor neutrinos, an~upper bound for the previous expression is~\cite{Balantekin:2006sw}
\begin{equation}
    \mu_\nu \geq (4\times 10^{-20})\mu_B.
\end{equation}

We, therefore, find that our estimate for the neutrino magnetic moment is more stringent than the SM~one.

\section{Summary and~Conclusions}

In this work, we study the neutrino chirality flip during the birth of an SN, produced by the possible existence of a neutrino magnetic moment. The~flip is caused by the interaction of the neutrino with electrons of the medium in the core of a magnetized strange quark matter NS. The~calculation is performed using average values of NS mass, radii, magnetic field intensities, temperature, and chemical potentials at the birth of NS. We compute the kick velocity induced by this anisotropic right-handed neutrino emission. By also~resorting to average values of the known kick velocities, we estimate the value $\mu_\nu \sim 3.6 \times 10^{-18}\mu_B$ for the neutrino magnetic moment. This estimate is more stringent than the SM bound obtained from bounds on the neutrino mass from solar, atmospheric, and reactor neutrinos. 
{We point out that the mechanism we hereby put forward does not discard other mechanisms as viable explanations for the observed kick velocities. In~fact, it may happen that different mechanisms coexist and are more or less important, depending on the given range of parameters describing NS, such as the mass, radius, temperature, and field strength. However, we also believe that the mechanism we put forward in this work can be at play, requiring only  a relatively small neutrino magnetic moment. In~this sense, the~work is also a motivation to continue searching for such a neutrino property. Also, notice that in the calculation, the assumption is implicit that the angle between the pulsar angular velocity and the magnetic field is small. This seems to be supported by the findings in Refs.~\cite{Young:2009iy,2021NatAs...5..788Y,Noutsos:2012dt}.}
We also point out that the inclusion of a neutrino magnetic moment and, thus, the possibility of chirality flip within the first seconds of the evolution of NS can have also other important consequences that should be explored in order to have a better characterization of these~objects.
\\

\section*{Acknowledgements}

Support for this work was received in part by UNAM-PAPIIT grant number IG100322 and by Consejo Nacional de Humanidades, Ciencia y Tecnolog\'ia grant numbers CF-2023-G-433. S. B. L. acknowledges the financial support of a fellowship granted by Consejo Nacional de Humanidades, Ciencia y Tecnolog\'ia as part of the Sistema Nacional de Posgrados.
%


%


\bibliography{biblio}

\end{document}